\documentclass[letterpaper,11pt]{article}

\usepackage[T1]{fontenc}
\usepackage[letterpaper,top=1in,bottom=1in,left=1in,right=1in]{geometry}

\usepackage[cmex10]{amsmath}
\usepackage{amsthm}
\usepackage{amssymb,latexsym}
\usepackage[usenames]{color}
\usepackage{graphicx}
\usepackage{psfrag}
\usepackage{xspace}
\usepackage{comment}
\usepackage{multirow}

\usepackage{comment}
\usepackage{xspace}

\newcommand{\sep}{,\xspace}

\newcommand{\incircle}{\textsf{Incircle}\xspace}

\newcommand{\vor}{Voronoi\xspace}
\newcommand{\rfl}{\mathcal{R}\xspace}
\newcommand{\rfx}[1]{\mathcal{R}(#1)\xspace}
\newcommand{\EE}{\mathbb{E}\xspace}

\newcommand{\ppp}{$PPP$\xspace}
\newcommand{\pps}{$PPS$\xspace}
\newcommand{\pss}{$PSS$\xspace}
\newcommand{\sss}{$SSS$\xspace}

\newcommand{\pppp}{$PPPP$\xspace}
\newcommand{\ppsp}{$PPSP$\xspace}
\newcommand{\pssp}{$PSSP$\xspace}
\newcommand{\sssp}{$SSSP$\xspace}

\newcommand{\ppps}{$PPPS$\xspace}
\newcommand{\ppss}{$PPSS$\xspace}
\newcommand{\psss}{$PSSS$\xspace}
\newcommand{\ssss}{$SSSS$\xspace}

    {\end{smallmatrix}\right|}

\newtheorem{theorem}{Theorem}

\newtheorem{lemma}[theorem]{Lemma}

\begin{document}
%
\title{Analysis of the \textsf{Incircle} predicate for the Euclidean\\
  Voronoi diagram of axes-aligned line segments}


\author{Manos N. Kamarianakis$^{1,2}$\hfil{}
Menelaos I. Karavelas$^{2,3}$\\[5pt]
\it{}$^1$Interdisciplinary Graduate Program
\it{}``Mathematics and their Applications''\\
\it{}University of Crete,
\it{}GR-714 09 Heraklion, Greece\\[5pt]
\it{}$^2$Department of Applied Mathematics,
\it{}University of Crete\\
\it{}GR-714 09 Heraklion, Greece\\[5pt]
\it{}$^3$Institute of Applied and Computational Mathematics\\
\it{}Foundation for Research and Technology - Hellas\\
\it{}P.O. Box 1385, GR-711 10 Heraklion, Greece\\[5pt]
{\texttt{mkaravel@tem.uoc.gr, manos@tem.uoc.gr}}\\[10pt]}

\maketitle

\begin{abstract}
  In this paper we study the most-demanding predicate for computing
the Euclidean Voronoi diagram of axes-aligned line segments, namely
the \textsf{Incircle} predicate. Our contribution
is two-fold: firstly, we describe, in algorithmic terms, how to
compute the \textsf{Incircle} predicate for axes-aligned line
segments, and  secondly we compute its algebraic degree.
Our primary aim is to minimize the algebraic degree, while, at the
same time, taking into account the amount of operations needed to
compute our predicate of interest.

In our predicate analysis we show that the \textsf{Incircle}
predicate can be answered by evaluating the signs of algebraic
expressions of degree at most 6; this is half the algebraic degree
we get when we evaluate the \textsf{Incircle} predicate using the
current state-of-the-art approach. In the most demanding cases
of our predicate evaluation, we reduce the problem of answering the
\textsf{Incircle} predicate to the problem of computing the sign of
the value of a linear polynomial (in one variable), when evaluated
at a known specific root of a quadratic polynomial (again in one
variable). Another important aspect of our approach is that, from a
geometric point of view, we answer the most difficult case of the
predicate via implicitly performing point locations on an
appropriately defined subdivision of the place induced by the
Voronoi circle implicated in the \textsf{Incircle} predicate.

  \bigskip\noindent
  \textit{Key\;words:}\/
  \incircle predicate\sep Euclidean Voronoi diagram\sep
  line segments\sep axes-aligned
\end{abstract}

\clearpage

\section{Introduction}

The Euclidean Voronoi diagrams of a set of line segments is one of the
most well studied structures in computational geometry. There are
numerous algorithms for its computation
\cite{dl-gvdp-78,k-eccs-79,l-matps-82,y-oavds-87,f-savd-87,
bdsty-arsol-92,kmm-ricav-93a}.
These include worst-case optimal algorithms that use different algorithmic
paradigms, such as the divide-and-conquer paradigm \cite{y-oavds-87} or the
sweep-line paradigm \cite{f-savd-87}. An interesting and efficient class of
algorithms rely on the randomized incremental construction of the
Voronoi diagram \cite{bdsty-arsol-92,kmm-ricav-93a}. From the
implementation point of view, there are algorithms that assume that
numerical computations are performed exactly \cite{s-avd-,k-reisv-04},
i.e., they follow the Exact Geometric Computation (EGC) paradigm
\cite{yd-ecp-95}, as well as algorithms that use floating-point
arithmetic \cite{i-toavd-96,siii-toiar-00,h-veare-01}; the latter 
class of algorithms does not guarantee exactness, but rather
topological correctness, meaning that the output of the algorithm has
the correct topology of a Voronoi diagram.
In terms of applications, these include computer graphics, pattern
recognition, mesh generation, NC machining and geographical
information systems (GIS) --- see
\cite{k-eccs-79,l-matps-82,by-ag-98,h-veare-01,g-dcssg-10}, and the
references therein.

Efficient and exact predicate evaluation in geometric algorithms is of
vital importance. It has to be fast for the algorithm to be efficient.
It has to be complete in the sense that it has to cover all degenerate
cases, which, despite that fact that they are ``degenerate'' from the
theoretical/analysis point-of-view, they are commonplace in real world
input. In the EGC paradigm context, exactness is the bare minimum that
is required in order to guarantee the correctness of the
algorithm. The efficiency of predicates is typically measured in terms
of the algebraic degree of the expressions (in the input parameters)
that are computed during the predicate evaluation, as well as the
number (and possibly type) of arithmetic operations involved. The goal
is not only to minimize the number of operations, but also
to minimize the algebraic degree of the predicates, since it is the
algebraic degree that determines the precision required for exact
arithmetic. Degree-driven approaches for either the evaluation of
predicates, or the design of the algorithm as a whole, has become an
important question in algorithm/predicate design over the past few
years \cite{b-ecvdl-96,LiPrTa99,bp-rpsis-00,DFMT02,KE03,EK06,ms-cpvdd-10}.

In this paper we are interested in the most demanding predicate of the
Euclidean Voronoi diagram of axes-aligned line segments. Axes-aligned
line segments, or line segments forming a 45-degree angle with respect
to the axes, are typical input instances in various applications, such
as VLSI design \cite{p-cacmm-01,gp-yao-08}. However, although the
predicates for the Euclidean \vor diagram of line segments have
already been studied \cite{b-ecvdl-96}, the predicates for
axes-aligned or ortho-45$^\circ$ input instances have not been studied
in detail in the Euclidean setting.
In the sections that follow, we analyze the \incircle predicate in this
setting: given three sites $S_1$, $S_2$ and $S_3$, such that the \vor
circle $V(S_1,S_2,S_3)$ exists, and a query object $O$, we seek to
determine if $O$ intersects the disk $D$ bounded by $V(S_1,S_2,S_3)$,
touches $D$ or is completely disjoint from $D$. In our context $S_1$,
$S_2$, $S_3$ and $O$ are either points or axes-aligned (open) line
segments.
Our aim is to minimize the algebraic degree of the expressions involved in
evaluating the \incircle predicate. We show that we can answer the
\incircle predicate using polynomial expressions in the input
quantities whose algebraic degree is at most 6. This is to be compared:
(1) against the generic bound on the maximum algebraic degree needed to
compute the \incircle predicate, when we impose no restriction on the
geometry of the line segments, which is 40 \cite{b-ecvdl-96}, and
(2) against the specialization/simplification of the approach in
\cite{b-ecvdl-96}, when we consider axes-aligned line segments. With
respect to the latter case, our algebraic degrees are never worse,
while in the most demanding case we have reduced the degree by a
factor of two (see also Table \ref{tbl:algdeg}).

The rest of our paper is structured as follows. In Section
\ref{sec:prelims} we give some definitions, compare our approach to
that in \cite{b-ecvdl-96}, and detail some of the tools that we use
in the \incircle predicate analysis. In Sections \ref{sec:ppp}-\ref{sec:pss}
we describe how we evaluate the incircle predicate for different
configurations of the sites $S_1$, $S_2$, $S_3$ and $O$. In Section
\ref{sec:concl} we detail plans for future work.

\section{Definitions and preliminaries}\label{sec:prelims}

Given three sites $S_1$, $S_2$, and $S_3$ we denote their \vor circle
by $V(S_1,S_2,S_3)$ (if it exists). There are at most two \vor circles
defined by the triplet $(S_1,S_2,S_3)$; the notation $V(S_1,S_2,S_3)$
refers to the \vor circle that ``discovers'' the sites $S_1$, $S_2$
and $S_3$ in that (cyclic) order, when we walk on the circle's
boundary in the counterclockwise sense. Given a fourth object $O$, which
we call the \emph{query object}, the \incircle predicate
$\incircle(S_1,S_2,S_3,O)$ determines the relative position $O$ with
respect to the disk $D$ bounded by $V(S_1,S_2,S_3)$. The predicate is
positive if $O$ does not intersect $D$, zero if $O$ touches the
boundary but not the interior of $D$, and negative of the
intersection of $O$ with the interior of $D$ is non-empty. 

The \vor circle of three sites does not always exist. In
this paper, however, we assume that the \incircle predicate is called
during the execution of an incremental algorithm for computing the
Euclidean Voronoi diagram of line segments, and thus the first three
sites are always sites related to a \vor vertex in the diagram.
Note that the value of the \incircle predicate does not change when we
circularly rotate the first three arguments. In that respect, there
are only four possible distinct configurations for the type of the \vor
circle: \ppp, \pps, \pss and \sss, where $P$ stands for point and $S$
stands for segment. For example, a \vor circle of \pss type goes
through a point and is tangent to two segments. This gives eight
possible configurations for the \incircle predicate, two per \vor circle
type.

The predicates for the Euclidean \vor diagram of line segments, in the
context of an incremental construction of the diagram, have already
been studied by Burnikel \cite{b-ecvdl-96}. According to Burnikel's
analysis the most demanding predicate is the \incircle
predicate: assuming that the input is either rational points, or
segments described by their endpoints as rational points, Burnikel
shows that the \incircle predicate can be evaluated using polynomial
expressions in the input quantities, whose algebraic degree is at most
40; this happens when the \vor circle is of \sss type and the query
object is a segment (see also the line dubbed ``General
\cite{b-ecvdl-96}'' in Table \ref{tbl:algdeg}). Considering Burnikel's
approach for the case of axes-aligned line segments, and performing
the appropriate simplifications in his calculations, we arrive at a
new set of algebraic degrees for the various configurations of the
\incircle predicate (see line dubbed ``Axes-aligned
\cite{b-ecvdl-96}'' in Table \ref{tbl:algdeg}); now the most demanding
case the is \pps case, which gives algebraic degree 8 and 12, when the
query object is a point and a segment, respectively. 

\begin{table}[t]
  \begin{center}
    \begin{tabular}{|c|c|c|c|c|}\hline
      &\pppp & \ppsp & \pssp & \sssp
      \\\hline\hline
      General \cite{b-ecvdl-96}&4&12&16&32\\
      Axes-aligned \cite{b-ecvdl-96}&4&8&4&2\\
      Axes-aligned [this paper]&4&6&4&2\\\hline
    \end{tabular}\\[5pt]
    \begin{tabular}{|c|c|c|c|c|}\hline
      &\ppps & \ppss & \psss & \ssss
      \\\hline\hline
      General \cite{b-ecvdl-96}&8&24&32&40\\
      Axes-aligned \cite{b-ecvdl-96}&6&12&4&2\\
      Axes-aligned [this paper]&6&6&4&2\\\hline
    \end{tabular}
  \end{center}
  \caption{Maximum algebraic degrees for the eight types of the
    \incircle predicate according to: \cite{b-ecvdl-96} for the general
    and the axes-aligned segments case, and
    this paper. Top/Bottom table: the query object is a
    point/segment.}
  \label{tbl:algdeg}
\end{table}

In Sections \ref{sec:ppp}-\ref{sec:pss} we analyze, in more or less detail,
all eight possible configurations for the \incircle predicate, and
show how we can reduce the algebraic degrees for the \pps case from 8
and 12 to 6. This is done by means of three key ingredients:
(1) we formulate the \incircle predicate as an algebraic problem of
the following form: we compute a linear polynomial $L(x)=l_1x+l_0$ and
a quadratic polynomial $Q(x)=q_2x^2+q_1x+q_0$, such that the result of
the \incircle predicate is the sign of $L(x)$ evaluated at a specific
root of $Q(x)$,
(2) for the \pps and \pss cases, we express the \incircle
predicate as a difference of distances, instead of as a difference of
squares of distances, and
(3) we reduce the \ppsp case to the \ppps case.
Regarding the first ingredient, we describe in the following
subsection how we can do better than finding the appropriate root of
$Q(x)$ and substitute it in $L(x)$ (this is essentially what is done
in \cite{b-ecvdl-96}). Regarding the second and third ingredients we
postpone the discussion until the corresponding sections.
There is one final tool that we will be very useful in order to
simplify our analysis: in order to reduce our case analysis we make
extensive use of the reflection transformation through the line
$y=x$; see Subsection \ref{sec:reflection} for the details.

\subsection{Evaluation of the sign of $L(x)=l_1x+l_0$ at a specific
  root of $Q(x)=q_2x^2+q_1x+q_0$}\label{sec:lqsolve}

Let $L(x)=l_1x+l_0$ and $Q(x)=q_2x^2+q_1x+q_0$ be a linear and a
quadratic polynomial, respectively, such that $Q(x)$ has non-negative
discriminant. Let the algebraic degrees of $l_1$, $l_2$, $q_2$, $q_1$
and $q_0$ be $\delta_l$, $\delta_l+1$, $\delta_q$, $\delta_q+1$, and
$\delta_q+2$, respectively. We are interested in the sign of
$L(r)$, where $r$ is one of the two roots $x_1\leq{}x_2$
of $Q(x)$. In our analysis below we will assume, without loss of
generality that $l_1,q_2>0$.

The obvious approach is to solve for $r$ and substitute into the
equation of $L(x)$. Let $ \Delta_Q=q_1^2-4q_2q_0$ be the discriminant of 
$Q(x)$. Then $r=(-q_1\pm\sqrt{\Delta_Q})/(2q_2)$, which in turn
yields $L(r)=(l_1q_1+2l_0q_2\pm\sqrt{\Delta_Q})/(2q_2)$. Computing
the sign of $L(r)$ is dominated, with respect to the algebraic
degree of the quantities involved, by the computation of the 
sign of $l_1q_1+2l_0q_2\pm{}l_1\sqrt{\Delta_Q}$. Evaluating the sign
of this quantity amounts to evaluating the sign of
$(l_1q_1+2l_0q_2)^2-l_1^2\Delta_Q$, which is of algebraic degree
$2(\delta_l+\delta_q+1)$.

Observe now that evaluating the sign of $L(r)$ is equivalent to
evaluating the sign of $Q(x^\star)$, and possibly the sign of
$Q'(x^\star)$, where $x^\star=-\frac{l_0}{l_1}$ stands for the root of
$L(x)$.
Indeed, if $Q(x^\star)<0$, we immediately know that $L(r)<0$
if $r\equiv{}x_1$, or that $L(r)>0$ if $r\equiv{}x_2$. If
$Q(x^\star)>0$, we need to additionally evaluate the sign of
$Q'(x^\star)=2q_2x^\star+q_1$. If $Q'(x^\star)<0$, we know that
$x^\star<x_1,x_2$, which implies that $L(r)>0$, whereas if
$Q'(x^\star)>0$, we have $x^\star>x_1,x_2$, which gives
$L(r)<0$. Finally, if $Q(x^\star)=0$, we still need to evaluate the
sign of $Q'(x^\star)$. If $Q'(x^\star)<0$, then $x^\star\equiv{}x_1$,
and thus $L(r)=0$ if $r\equiv{}x_1$, and $L(r)>0$ if
$r\equiv{}x_2$. Similarly, if $Q'(x^\star)>0$, then $x^\star\equiv{}x_2$,
and thus $L(r)<0$ if $r\equiv{}x_1$, and $L(r)=0$ if
$r\equiv{}x_2$. There is one last case to consider:
$Q'(x^\star)=0$. Given that $Q(x^\star)=0$, this can happen only if
$x_1=x_2=x^\star$, in which case we deduce $L(r)=0$.
Since $Q(x^\star)=(l_1^2q_0-l_1q_1l_0+q_2l_0^2)/l_1^2$, evaluating
the sign of $Q(x^\star)$ means evaluating the sign of an algebraic
expression of degree $2\delta_l+\delta_q+2$. Moreover,
$Q'(x^\star)=(l_1q_1-2q_2l_0)/l_1$; hence, evaluating the sign of
$Q'(x^\star)$ reduces to evaluating the signs of $l_1q_1-2q_2l_0$ and
$l_1$, the degrees of which are $\delta_l+\delta_q+1$ and $\delta_l$,
respectively. 

Notice that the latter among the two approaches described above is never
worse than the first one; in fact, if $\delta_q>0$ it gives a lower
maximum algebraic degree. We summarize this observation in the
following lemma.

\begin{lemma}\label{lemma:lqsolve}
  Let $L(x)=l_1x+l_0$, $l_1\ne{}0$, and $Q(x)=q_2x^2+q_1x+q_0$,
  $q_2\ne{}0$, be a linear and quadratic polynomial, respectively,
  such that the discriminant of $Q(x)$ is non-negative. If the
  algebraic degrees of $l_1$, $l_2$, $q_2$, $q_1$ and $q_0$ be
  $\delta_l$, $\delta_l+1$, $\delta_q$, $\delta_q+1$, and
  $\delta_q+2$, respectively, then we can evaluate the sign of
  $L(r)$, where $r$ is a specific root of $Q(x)$, using
  expressions of maximum algebraic degree $2\delta_l+\delta_q+2$.
\end{lemma}

\subsection{Reflection transformation}
\label{sec:reflection}

\begin{figure}[!h]
  \begin{center}
    \includegraphics[width=0.7\textwidth]{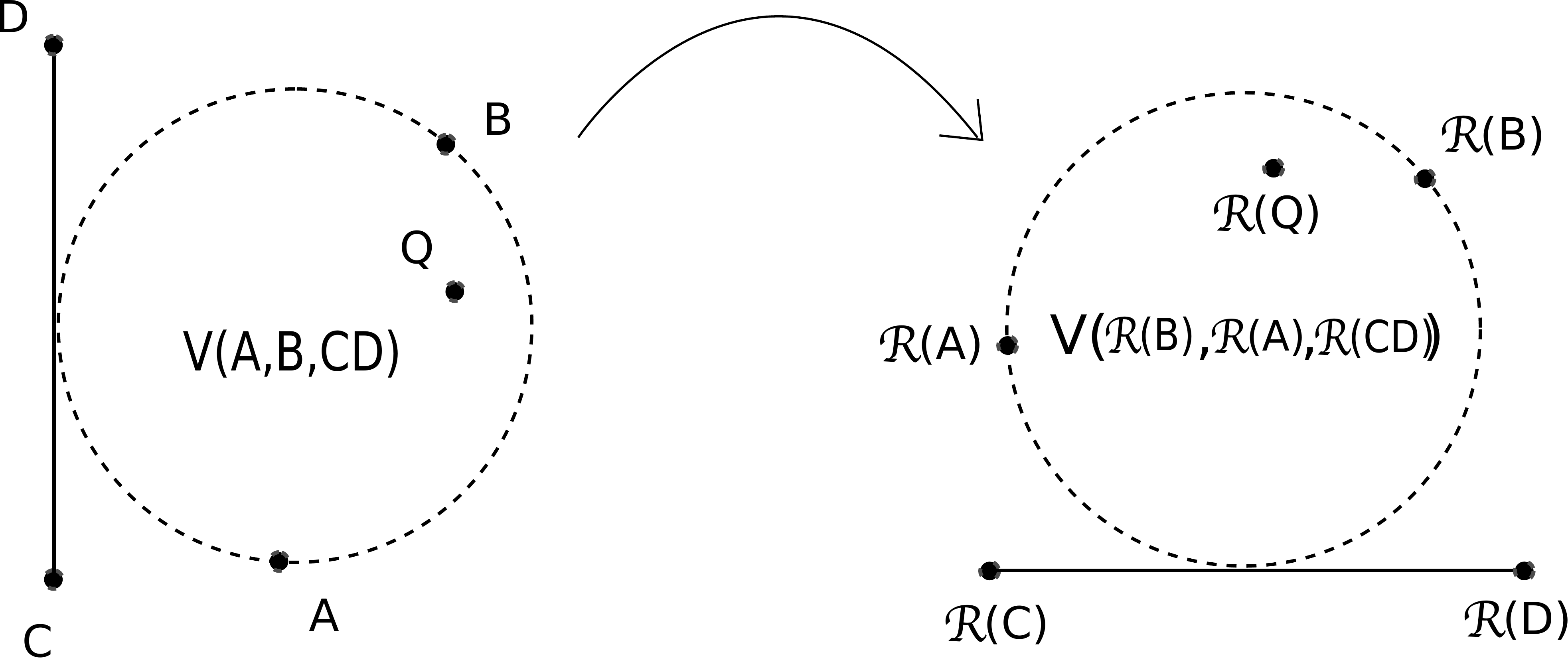} 
  \end{center}
  \caption{$\incircle(A,B,CD,Q)$ is equivalent to
    $\incircle(\rfx{B},\rfx{A},\rfx{CD},\rfx{Q})$, where
    $\rfl$ stands for the image of $I$ under the reflection
    transformation through the line $y=x$.}
  \label{fig:transform}
\end{figure}

Let $\rfl:\EE^2\rightarrow\EE^2$ denote the reflection transformation
through the line $y=x$. $\rfl$ maps a point $(x,y)\in\EE^2$ to the
point $(y,x)\in\EE^2$. 
The reflection transformation preserves circles and line segments and
is inclusion preserving. This immediately implies that, given a \vor
circle $V(S_1,S_2,S_3)$ defined by three sites $S_1$, $S_2$ and $S_3$,
and a query point $Q$, $Q$ lies inside, on, or outside the \vor circle
$V(S_1,S_2,S_3)$ if and only if $\rfx{Q}$ lies inside, on, or outside
the \vor circle $V(\rfx{S_2},\rfx{S_1},\rfx{S_3})$ (cf. Fig. 
\ref{fig:transform} for the case where $S_1$ and $S_2$ are points and
$S_3$ is a segment). Hence,
$\incircle(S_1,S_2,S_3,Q)=\incircle(\rfx{S_2},\rfx{S_1},\rfx{S_3},\rfx{Q})$.
Notice that reflection reverses the orientation of a circle, which is
why we consider the \vor circle $V(\rfx{S_2},\rfx{S_1},\rfx{S_3})$
instead of the \vor circle $V(\rfx{S_1},\rfx{S_2},\rfx{S_3})$.
The same principle applies in the case where the query object is a
line segment $QS$:
$\incircle(S_1,S_2,S_3,QS)=\incircle(\rfx{S_2},\rfx{S_1},\rfx{S_3},\rfx{QS})$.

As a final note, the reflection transformation $\rfl$ maps an $x$-axis
parallel segment to a $y$-axis parallel segment, and vice versa. This
property will be used, in the sections that follow, to reduce the
analysis and computation of the \incircle predicate, where one of the
$S_i$'s is $y$-axis parallel, to the case where one of the $S_i$'s is
$x$-axis parallel.

\section{The \ppp case}\label{sec:ppp}

As of this section, we discuss and analyze the \incircle predicate for
each of the four possible configurations for the \vor circle. We start
with the case where the \vor circle is defined by three points $A$,
$B$ and $C$.

\subsection{The query object is a point}\label{sec:pppp}
This is the well known \incircle predicate for four points $A$, $B$,
$C$ and $Q$, where $Q$ is the query point, and it amounts to the
computation of the sign of the determinant
\[
\incircle(A,B,C,Q) = \begin{vmatrix}
1 & x_{A} & y_{A} & x_{A}^{2} + y_{A}^{2} \\
1 & x_{B} & y_{B} & x_{B}^{2} + y_{B}^{2} \\
1 & x_{C} & y_{C} & x_{C}^{2} + y_{C}^{2} \\
1 & x_{Q} & y_{Q} & x_{Q}^{2} + y_{Q}^{2} 
\end{vmatrix}.
\]
Its algebraic degree is clearly 4. 


\subsection{The query object is a segment}\label{sec:ppps}

Let $QS$ be the query segment.
In this case, we must first check that relative position of $Q$ and
$S$ with respect to $V(A,B,C)$ using $\incircle(A,B,C,I)$, $I\in\{
Q,S\}$. If at least one of $Q$ and $S$ lies inside $V(A,B,C)$, we
clearly have $\incircle(A,B,C,QS)<0$. 

Otherwise, we have to examine if the segment $QS$ intersects with
$V(A,B,C)$. This is equivalent to point-locating the points $Q$ and
$S$ in the arrangement of the lines $y=y_{min}$, $y=y_{max}$ and
$x=x_K$ if $QS$ is $x$-axis parallel or, $x=x_{min}$, $x=x_{max}$ and $y=y_K$ if 
$QS$ is $y$-axis parallel, where $x_{min}$, $x_{max}$ (resp.,
$y_{min}$ ,$y_{max}$) are the extremal points of $V(A,B,C)$ in the
direction of the $x$-axis (resp., $y$-axis). In fact, the case where
the segment $QS$ is $y$-axis parallel can be reduced to the case where
the query segment is $x$-axis parallel by noting that 
$\incircle(A,B,C,QS)=\incircle(\rfx{B},\rfx{A},\rfx{C},\rfx{QS})$
(see Section \ref{sec:reflection}). We will
therefore restrict our analysis to the case where $QS$ is $x$-axis
parallel. 

\begin{figure}[b]
  \begin{center}
    \includegraphics[width=0.7\textwidth]{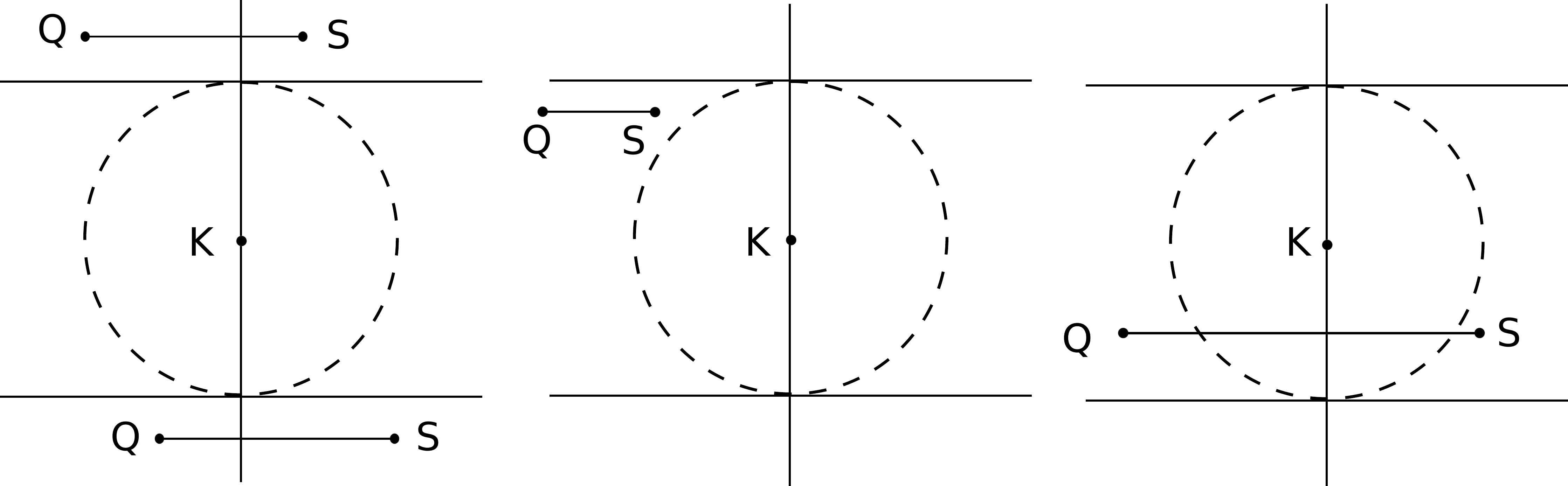} 
  \end{center}
  \caption{Relative positions of the $x$-axis aligned query segment
    $QS$ with respect to the lines $x=x_K$, $y=y_{min}$, $y=y_{max}$.}
  \label{fig:ppps}
\end{figure}

We first determine if $Q$ lies outside the band delimited by the lines
$y=y_{min}$ and $y=y_{max}$; in this case we immediately get
$\incircle(A,B,C,QS) >0$. Otherwise, if $Q$ lies inside the band
(resp., $Q$ lies on either $y=y_{min}$ or $y=y_{max}$), we check the
relative positions of $Q$ and $S$ against the line $x=x_K$; the
segment $QS$ intersects (resp., is tangent to) $V(A,B,C)$ if and only
if $Q$ and $S$ lie on different sides of the line $x=x_K$. 

In order to determine the relative position of $Q$ with respect to the lines 
$y=y_{min}$ and $y=y_{max}$, we will evaluate a quadratic $y$-polynomial 
that vanishes at $y_{min}$ and $y_{max}$: let $T(y)=t_2y^2+t_1y+t_0$
be this polynomial.
Having computed this polynomial, $y_Q\in(y_{min},y_{max})$ if and only
if $sign(T(y_Q)) =-sign(t_2) $,
$y_Q\not\in[y_{min},y_{max}]$ if and only if $sign(T(y_Q))=sign(t_2)$,
and, finally, $y_Q\in\{y_{min},y_{max}\}$ if and only if $sign(T(y_Q))=0$.

To evaluate such a polynomial, we first observe that every point $(x,y)$
on $V(A,B,C)$ satisfies $ \incircle(A,B,C,(x,y))=0$. 
Expanding the four-point \incircle determinant in terms of $x$, we end
up with a quadratic polynomial $U(x;y) = u_2x^2+u_1x+u_0(y)$ for
$\incircle(A,B,C,(x,y))$, where
\begin{equation*}
u_2 = \begin{vmatrix}
1 & x_A & y_A\\
1 & x_B & y_B\\
1 & x_C & y_C
\end{vmatrix}, \qquad
u_1 = \begin{vmatrix}
1 & y_A & x_A^2 +y_A^2\\
1 & y_B & x_B^2 +y_B^2\\
1 & y_C & x_C^2 +y_C^2
\end{vmatrix}, \qquad
u_0(y) = \begin{vmatrix}
1 & x_A & y_A & x_A^2 + y_A^2 \\
1 & x_B & y_B & x_B^2 + y_B^2 \\
1 & x_C & y_C & x_C^2 + y_C^2 \\
1 & 0 & y &   y^2
\end{vmatrix}
\end{equation*}

For a fixed value $y^\star$ of $y$, the roots of $U(x;y^\star)$
are the points of intersection of the line $y=y^\star$ with the \vor
circle $V(A,B,C)$. $U(x;y^\star)$ has no real roots if
$y^\star\not\in[y_{min},y_{max}]$, has two distinct roots if
$y^\star\in(y_{min},y_{max})$ and has a double root if
$y^\star\in\{y_{min},y_{max}\}$. In the last case, the discriminant
$\Delta_U(y^\star)=u_1^2-4u_2u_0(y^\star)$ of $U(x;y^\star)$ has to
vanish. Now consider the discriminant as a polynomial of $y$. Clearly,
$\Delta_U(y)$ is a quadratic $y$-polynomial, with a strictly
negative, since the points $A$, $B$ and $C$ are not
collinear. Moreover, $\Delta_U(y)$ vanishes for
$y\in\{y_{min},y_{max}\}$, hence it may serve as the quadratic
polynomial $T(y)$ we were aiming for.
More specifically, $T(y):=\Delta_U(y)=t_2y^2+t_1y+t_0$ where, 
$t_2 = -4 u_2^2$, $t_1 = 4 u_2 w_1$, $t_0 = u_1^2+4 u_2 u_3$, and
\begin{equation*}
w_1 = \begin{vmatrix}
1 & x_A & x_A^2 +y_A^2\\
1 & x_B & x_B^2 +y_B^2\\
1 & x_C & x_C^2 +y_C^2
\end{vmatrix}, \qquad
u_3 = \begin{vmatrix}
x_A & y_A & x_A^2 + y_A^2 \\
x_B & y_B & x_B^2 + y_B^2 \\
x_C & y_C & x_C^2 + y_C^2 
\end{vmatrix}
\end{equation*}

In an analogous manner, we can evaluate a quadratic $x$-polynomial
that vanishes at $x_{min}$ and $x_{max}$, which we call $S(x)$. More
precisely, $S(x) = s_2x^2+s_1x+s_0$, where
$s_2 = -4 u_2^2$, $s_1 = -4 u_2 u_1$ and $s_0 = w_1^2+4 u_2 u_3$.
In order to determine the relative position of  $Q$ and $S$ with
respect to the line $x=x_K$, we use the fact that
$x_K=\frac{1}{2}(x_{min}+x_{max})=-\frac{s_1}{s_2}$. Hence, using the
fact that $s_2<0$, checking on which side of $x=x_K$ lies point $I$,
for $I\in\{Q,S\}$, amounts to determining the sign
$sign(x_K-x_I)=sign(2s_2x_I+s_1)$.

The algebraic degrees of $u_0$, $u_1$, $u_2$, $u_3$, and $w_1$ are 4,
3, 2, 3, and 3, respectively. Therefore, the algebraic degrees of
$t_2$, $t_1$, $t_0$, $s_2$, $s_1$, and $s_0$ are 4, 5, 6, 4, 5, and 6,
respectively. This implies that the algebraic degree of $T(y_Q)$ is 6,
while the algebraic degree of $s_2x_I+s_1$, $I\in\{Q,S\}$, is 5. We,
thus, conclude that we can answer the \incircle predicate in the \ppps
case by evaluating expressions of maximum algebraic degree 6.


\section{The \sss case}\label{sec:sss}

In this section we consider the case where the \vor circle is defined
by three axis-aligned segments $AB$, $CD$ and $FG$. In order for the
circle $V(AB,CD,FG)$ to be well defined, exactly two of these segments
must parallel to each other, while the third perpendicular to the
other two.
Given that $V(AB,CD,FG)\equiv{}V(FG,AB,CD)\equiv{}V(CD,FG,AB)$, we
can assume without loss of generality that the first two segments are
parallel to each other, and thus the third is perpendicular to the
first two. 
Hence we only have to consider two cases:
(1) $AB$, $CD$ are $x$-axis parallel and $FG$ is $y$-axis parallel,
and
(2) $AB$, $CD$ are $y$-axis parallel and $FG$ is $x$-axis parallel.
In fact the second case can be reduced to the first one by noting that
$\incircle(AB,CD,FG,Q)=\incircle(\rfx{CD},\rfx{AB},\rfx{FG},\rfx{Q})$
(see Section \ref{sec:reflection}).
We shall, therefore, assume that $AB$, $CD$ are $x$-axis parallel and 
$FG$ is $y$-axis parallel.

\subsection{The query object is a point}\label{sec:sssp}
  
Let $Q$ be the query point. Since the center $K$ of $V(AB,CD,FG)$ lies
on the bisector of the lines $ \ell_{AB}$ and $\ell_{CD}$, and the
radius $\rho$ of the circle is the distance of $K$ from either
$\ell_{AB}$ or $\ell_{CD}$ (i.e., half the distance of the two lines),
we have 
\begin{equation}\label{equ:lll-vc}
  K= (x_F+\dfrac{y_C-y_A}{2}, \dfrac{y_C+y_A}{2}), \quad \rho =
  \dfrac{|y_C-y_A|}{2}.
\end{equation}

To answer the \incircle predicate for $Q$, we first examine if $Q$ and
$K$ lie on the same side with respect to the lines $\ell_{AB}$,
$\ell_{CD}$ and $\ell_{FG}$. If this is not the case, we immediately
conclude that $\incircle(AB,CD,FG,Q) >0$. Otherwise we must compare 
the distance $d(Q,K)$ of $Q$ from $K$ against the \vor radius
$\rho$. More precisely: $\incircle(AB,CD,FG,Q) = sign(d^2(Q,K)-\rho^2)$,
where $4(d^2(Q,K)-\rho^2)=4 (x_F-x_Q)(1+y_C-y_A)+(y_C+y_A-2y_Q)^2$,
which is an algebraic expression of degree 2 in the input
quantities. Given that the sideness tests for $Q$ against the lines
$\ell_{AB}$, $\ell_{CD}$ and $\ell_{FG}$ are of degree 1, we conclude
that answering the \incircle predicate in the \sssp case amounts to
computing the signs of expressions of algebraic degree at most 2.

\subsection{The query object is a segment}\label{sec:ssss}

Let $QS$ be the query segment. We first determine if the endpoints $Q$
and/or $S$ of $QS$ lie inside $V(AB,CD,FG)$, in which case we
immediately get $\incircle(AB,CD,FG,QS)<0$. Otherwise, we must
consider the orientation of $QS$ and make the appropriate checks.

Assume first that $QS$ is $x$-axis parallel. We first check if $Q$ is
inside the band $B_y$ delimited by the lines $\ell_{AB}$ and
$\ell_{CD}$. If $Q$ lies outside $B_y$, we immediately get that
$\incircle(AB,CD,FG,QS)>0$. Otherwise, we have to determine the
relative positions of $Q$ and $S$ with respect to the line $x=x_K$,
where $x_K=x_F+\frac{1}{2}(y_C-y_A)$, by evaluating the signs  
of $x_Q-x_K$ and $x_S-x_K$. If $Q$ lies inside $B_y$ (resp., on the
boundary  of $B_y$, $QS$ intersects (resp., is tangent to)
$V(AB,CD,FG)$, if and only if $Q$ and $S$ lie on different sides of
the line $x=x_K$, i.e., if and only if
$(x_Q-x_K)(x_S-x_K)<0$. Determining if $Q$ lies inside $B_y$ amounts
to computing the signs of $y_Q-y_A$ and $y_Q-y_C$, which are degree 1
quantities. The quantities $x_Q-x_K$ and $x_S-x_K$ are also of degree
1, which implies that we can answer the \incircle predicate in this
case using quantities of algebraic degree up to 2 (the algebraic
degree needed to evaluate $\incircle(AB,CD,FG,I)$, $I\in\{Q,S\}$
dominates the degrees of all other quantities to be evaluated).

Consider now the case where $QS$ is $y$-axis parallel. We first need
to check if the line $\ell_{QS}$, intersects with $V(AB,CD,FG)$. To do
this we need to evaluate the sign of quantity $|x_Q-x_K|-\rho$, where
$\rho$ is given by \eqref{equ:lll-vc}. Computing the signs of
$x_Q-x_K$ and $y_C-y_A$, we may express $|x_Q-x_K|-\rho$ as a
polynomial expression in the input quantities; its algebraic degree is,
clearly, 1. If $|x_Q-x_K|-\rho>0$, $\ell_{QS}$ does not intersect
$V(AB,CD,FG)$, and we immediately get $\incircle(AB,CD,FG,QS)>0$.
Otherwise, if $|x_Q-x_K|-\rho<0$ (resp., $|x_Q-x_K|-\rho=0$)
$\ell_{QS}$ either intersects with (resp. either is tangent to) the
\vor circle or does not intersect the \vor circle at all. To
distinguish between these two cases we have to determine if the points
$Q$ and $S$ lie on different sides of the line $y=y_K$: the segment
$QS$ intersects with (resp., is tangent to) the \vor circle
$V(AB,CD,FG)$ if and only if $(y_Q-y_K)(y_S-y_K)<0$. Since
$y_K=\frac{1}{2}(y_A+y_C)$ (see rel. \eqref{equ:lll-vc}), determining
the signs $sign(y_Q-y_K)$ and $sign(y_S-y_K)$ amounts to computing the
sign of quantities of algebraic degree 1. As in the case where $QS$ is
$x$-axis parallel, the algebraic degree for evaluating the \incircle
predicate is dominated by the algebraic degree for evaluating
$\incircle(AB,CD,FG,I)$, $I\in\{Q,S\}$, which is 2.


\section{The generic approach for the evaluation of the \incircle
  predicate in the \pps and \pss cases}\label{sec:generic}

In this section we present our approach for evaluating the
\incircle predicate in a generic manner. The approach presented is
applicable when the \vor circle is defined by at least one point and
at least one segment, i.e., we can treat the cases \pps and \pss.

Let $K=(x_K,y_K)$ be the center of the Voronoi circle defined by the sites
$S_1$, $S_2$, $S_3$, that touches the sites $S_1$, $S_2$ and $S_3$ in
that order when we traverse the \vor circle in the counterclockwise
sense.
As already stated, we want to evaluate the \incircle predicate for a
query point or a query line segment with respect to this circle.
To do this we compute a quadratic polynomial $P(x)$ that 
vanishes at $x_K$, while using geometric considerations and the
requirement on the orientation of the \vor circle, we can determine
which of the roots $x_1\leq{}x_2$ of $P(x)$ corresponds to $x_K$. 
Regarding $y_K$, the situation is entirely symmetric. We also
compute a quadratic polynomial $T(y)$ that vanishes at $y_K$ and, as
for $x_K$, we can determine which of the two roots $y_1\leq{}y_2$ of
$T(y)$ corresponds to $y_K$. Moreover, in all cases $x_K$ and $y_K$
are linearly dependent, which means that we may express $y_K$ as
$y_K=\frac{\alpha_1}{\beta}x_K+\frac{\alpha_0}{\beta}$, where $\alpha_1$,
$\alpha_0$ and $\beta$ are polynomials in the input quantities.

\subsection{The query site is a point}\label{sec:pxsp}

Let $Q$ be the query point. Since at least one of $S_1$, $S_2$ and
$S_3$ is a point $A$, determining the \incircle predicate amounts to
evaluating the sign of the quantity
$d^2(K,Q)-d^2(K,A)=(x_K-x_Q)^2+(y_K-y_Q)^2-(x_K-x_A)^2-(y_K-y_A)^2$.
Replacing $y_K$, using the relation
$y_K=\frac{\alpha_1}{\beta}x_K+\frac{\alpha_0}{\beta}$, and gathering
the terms of $x_K$, we get
$\incircle(S_1,S_2,S_3,Q)=\frac{1}{\beta}(I_1x_K+I_0)$, where
$I_1=2\beta(x_Q-x_A)+2\alpha_1(y_Q-y_A)$ and
$I_0=\beta(x_Q^2+y_Q^2-x_A^2-y_A^2)-2\alpha_0(y_Q-y_A)$.
If $I_1=0$, the we can immediately evaluate the \incircle predicate by
evaluating the signs of $I_0$ and $\beta$. Otherwise, deciding the
\incircle predicate reduces to evaluating the sign of $\beta$, as well
as the sign of $I_1x+I_0$, evaluated at a specific known root of a
quadratic polynomial $P(x)=p_2x^2+p_1x+p_0$ (it is the root of $P(x)$
that corresponds to $x_K$). This is exactly the problem we analyzed in
Subsection \ref{sec:lqsolve}.

Let us now analyze the algebraic degrees of the expressions above. As
we will see in the upcoming sections (see Sections \ref{sec:pps} and
\ref{sec:pss}), $P(x)$ is a homogeneous polynomial in terms of its
algebraic degree. Letting $\delta_x$ the algebraic degree of $p_2$,
the algebraic degrees of $p_1$ and $p_0$ become $\delta_x+1$ and
$\delta_x+2$. Let also $\delta_\alpha$ be the algebraic degree of
$\alpha_1$. In our context, the algebraic degree of $\alpha_0$ is
always one more that the degree of $\alpha_1$, i.e., it is
$\delta_\alpha+1$, whereas the algebraic degree of $\beta$ is always
equal to that of $\alpha_1$. This implies that the algebraic
degrees of $I_1$ and $I_0$ are $\delta_\alpha+1$ and
$\delta_\alpha+2$, respectively. Applying Lemma \ref{lemma:lqsolve}, we
conclude that we can resolve resolve the \incircle predicate using
expressions of maximum algebraic degree
$2(\delta_\alpha+1)+\delta_x+2=2\delta_\alpha+\delta_x+4$.

\subsection{The query site is a segment}\label{sec:pxss}

Let $QS$ be the query segment. The first step is to compute
$\incircle(S_1,S_2,S_3,Q)$ and, if needed,
$\incircle(S_1,S_2,S_3,S)$. If at least one $Q$ and $S$ lies inside
the \vor circle $V(S_1,S_2,S_3)$, we get
$\incircle(S_1,S_2,S_3,QS)<0$. Otherwise, we need to determine if the
line $\ell_{QS}$ intersects $V(S_1,S_2,S_3)$. If $\ell_{QS}$ does not
intersect the \vor circle, we have $\incircle(S_1,S_2,S_3,QS)>0$. If
$\ell_{QS}$ intersects the \vor circle we have to check if $Q$ and $S$
lie on the same or opposite sides of the line $\ell^\perp_{QS}(K)$ that
goes through the \vor center $K$ and is perpendicular to
$\ell_{QS}$. Notice that since $QS$ is axes-aligned, the line
$\ell^\perp_{QS}(K)$ is either the line $x=x_K$ or the line $y=y_K$.
Since at least one of $S_1$, $S_2$ and $S_3$ is a segment $CD$,
answering the \incircle predicate is equivalent to comparing the
distance of $K$ from the line $\ell_{QS}$ to the segment $CD$: 
\begin{equation}\label{equ:incircle-xxss}
  \incircle(S_1,S_2,S_3,\ell_{QS}) = d(K,\ell_{QS})-d(K,CD).
\end{equation}
We can assume without loss of generality that $CD$ is
$x$-axis parallel, since, otherwise we can reduce
$\incircle(S_1,S_2,S_3,QS)$ to
$\incircle(\rfx{S_2},\rfx{S_1},\rfx{S_3},$ $\rfx{QS})$ (see Section
\ref{sec:reflection}), in which case $\rfx{CD}$ is $x$-axis parallel.
Let us now examine and analyze the right-hand side difference of
\eqref{equ:incircle-xxss}.

Assume first that the segment $QS$ is $x$-axis parallel. In this case the
equation of $\ell_{QS}$ is $y=y_Q$, and, hence,
$d(K,\ell_{QS})=|y_K-y_Q|$. Recall that $y_K$ is a specific root of a
quadratic polynomial $T(y)$. Therefore, determining the sign of
$y_K-y_Q$ reduces to evaluating the sign of $T(y_Q)$ and $T'(y_Q)$.
Let $T(y)=t_2y^2+t_1y+t_0$ be this polynomial, and let $\delta_y$,
$\delta_y+1$, $\delta_y+2$ be the algebraic degrees of $t_2$, $t_1$ and
$t_0$, respectively (as for $P(x)$, $T(y)$ is a homogeneous polynomial).
Consider now the case where $QS$ is $y$-axis parallel. The equation of
$\ell_{QS}$ is $x=x_Q$, and, hence, $d(K,\ell_{QS})=|x_K-x_Q|$. As in
the $x$-axis parallel case, $x_K$ is a specific known root of the 
quadratic polynomial $P(x)$, and determining the sign of $x_K-x_Q$
amounts to evaluating the sign of $P(x_Q)$ and $P'(x_Q)$. 
%
%
Last but not least, since the segment $CD$ is $x$-axis parallel,
$d(K,CD)=|y_K-y_C|$. As before, we can determine the sign of $y_K-y_C$
by evaluating the signs of $T(y_C)$ and $T'(y_C)$.

Having made the above observations, we conclude that, if $QS$ is
$x$-axis parallel,
\[\incircle(S_1,S_2,S_3,\ell_{QS})=|y_K-y_Q|-|y_K-y_C|=J_1y_K+J_0,\]
where $J_1$ and $J_0$ are given in the following table.
\begin{center}
  \begin{tabular}{|c|c|c|c|}\hline 
    $y_K-y_Q$&$y_K-y_C$&$J_1$&$J_0$\\\hline\hline
    \multirow{2}{*}{$\geq 0$}&$\geq{}0$&$0$&$y_C-y_Q$\\\cline{2-4}
    &$<0$&$2$&$-y_Q-y_C$\\\hline
    \multirow{2}{*}{$<0$}&$\geq{}0$&$-2$&$y_Q+y_C$\\\cline{2-4}
    &$<0$&$0$&$-y_C+y_Q$\\\hline 
  \end{tabular}
\end{center}
Clearly, if $J_1=0$ we have
$\incircle(S_1,S_2,S_3,\ell_{QS})=sign(J_0)$. Otherwise,
given that $y_K$ is a root of $T(y)$, evaluating
$\incircle(S_1,S_2,S_3,\ell_{QS})$ can be done using the analysis in
Subsection \ref{sec:lqsolve}. Since the algebraic degrees of $J_1$ and
$J_0$ are 0 and 1, respectively, we deduce, by Lemma
\ref{lemma:lqsolve}, that we can resolve the \incircle predicate using
expressions of algebraic degree at most
$2\cdot{}0+\delta_y+2=\delta_y+2$.

For the case where $QS$ is $y$-parallel we use the fact that
$y_K=\frac{\alpha_1}{\beta}x_K+\frac{\alpha_0}{\beta}$. Using this
linear dependence between $x_K$ and $y_K$, we get
\[\incircle(S_1,S_2,S_3,\ell_{QS})=|x_K-x_Q|-|y_K-y_C|
=\frac{1}{\beta}(L_1x_K+L_0),\]
where $L_1$ and $L_0$ are given in the following table.
\begin{center}
  \begin{tabular}{|c|c|c|c|}\hline 
    $x_K-x_Q$&$y_K-y_C$&$L_1$&$L_0$\\\hline\hline
    \multirow{2}{*}{$\geq 0$}
    &$\geq{}0$&$-\alpha_1+\beta$&$\beta(y_C-x_Q)-\alpha_0$\\\cline{2-4}
    &$<0$&$\alpha_1+\beta$&$\beta(-y_C-x_Q)+\alpha_0$\\\hline
    \multirow{2}{*}{$<0$}
    &$\geq{}0$&$-\alpha_1-\beta$&$\beta(y_C+x_Q)-\alpha_0$\\\cline{2-4}
    &$<0$&$\alpha_1-\beta$&$\beta(-y_C+x_Q)+\alpha_0$\\\hline 
  \end{tabular}
\end{center}
If $L_1=0$,
$\incircle(S_1,S_2,S_3,\ell_{QS})=sign(L_0)sign(\beta)$.
Otherwise, given that $x_K$ is a known root of $P(x)$,
determining the sign of $L_1x_K+L_0$ can be done as in Subsection
\ref{sec:lqsolve}. As in the previous subsection, we let
$\delta_\alpha$ be the algebraic degree of $\alpha_1$ (and also of
$\beta$), which means that the degree of $\alpha_0$ is
$\delta_\alpha+1$. Hence, the algebraic degree of $L_1$ is
$\delta_\alpha$, whereas that of $L_0$ is
$\max\{\delta_\alpha+1,1\}=\delta_\alpha+1$. By Lemma
\ref{lemma:lqsolve}, in order to evaluate the sign $L_1x_K+L_0$ we
need to compute the signs of expressions of algebraic degree at most
$2\delta_\alpha+\delta_x+2$.

As we mentioned at the beginning of this subsection, if
$\incircle(S_1,S_2,S_3,\ell_{QS})\le{}0$, we need to check the
position of $Q$ and $S$ with respect to the either line $x=x_K$ (if
$QS$ is $x$-axis parallel), or the line $y=y_K$ (if $QS$ is $y$-axis
parallel). To check the position of $I$, $I\in\{Q,S\}$, against the
line $x=x_K$, we simply have to compute the signs of $P(x_I)$ and
$P'(x_I)$. The algebraic degrees of these quantities are $\delta_x+2$
and $\delta_x+1$, respectively. In a symmetric manner, to check the
position of $I$, $I\in\{Q,S\}$, against the line $y=y_K$, we
simply have to compute the signs of $T(y_I)$ and $T'(y_I)$. The
algebraic degrees of these quantities are $\delta_y+2$ and
$\delta_y+1$, respectively. Notice that in both cases for the
orientation of $QS$, the algebraic degree of the quantities whose sign
needs to be evaluated to resolve the \incircle predicate are never
greater than those computed above for evaluating
$\incircle(S_1,S_2,S_3,\ell_{QS})$.
Recalling that, in order to evaluate $\incircle(S_1,S_2,S_3,QS)$, the
first step is to evaluate $\incircle(S_1,S_2,S_3,Q)$, and, if needed,
$\incircle(S_1,S_2,S_3,S)$, we conclude that in order to evaluate the
\incircle predicate when the query object is a segment we need to
compute the sign of polynomial expressions of algebraic degree
at most $\max\{2\delta_\alpha+\delta_x+4,\delta_y+2\}$.

\section{The \pps case}\label{sec:pps}

Let $A$ and $B$ be the two points and $CD$ be the segment defining the
\vor circle. Without loss of generality we may assume that $CD$ is
$x$-axis parallel, since otherwise we can reduce $\incircle(A,B,CD,Q)$
to $\incircle(\rfx{B},\rfx{A},\rfx{CD},\rfx{Q})$, as described in
Section \ref{sec:reflection}.

\subsection{The query object is a point}\label{sec:ppsp}

\begin{figure}[!b]
  \begin{center}
    \includegraphics[width=0.4\textwidth]{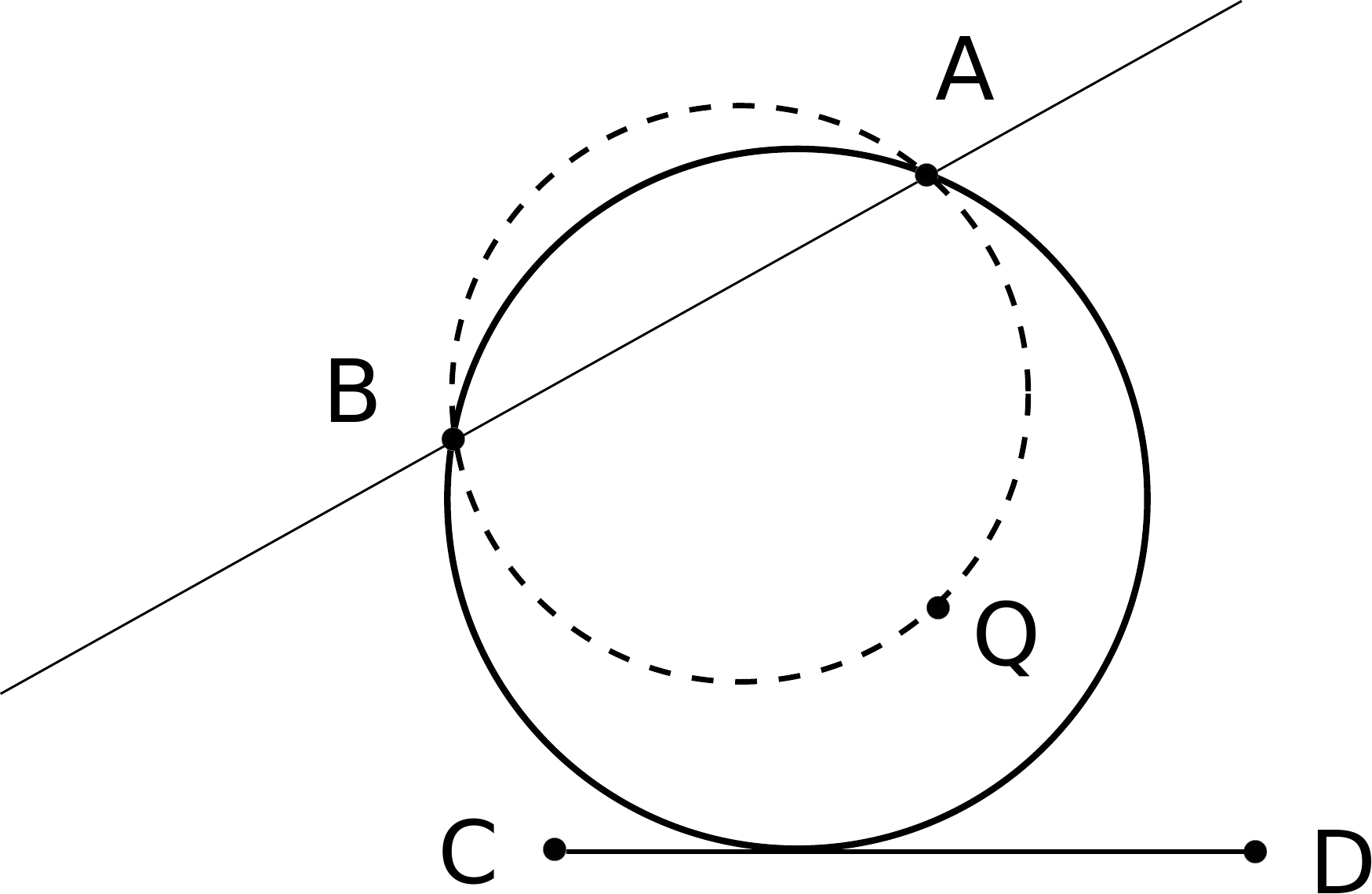}\hfil%
    \includegraphics[width=0.4\textwidth]{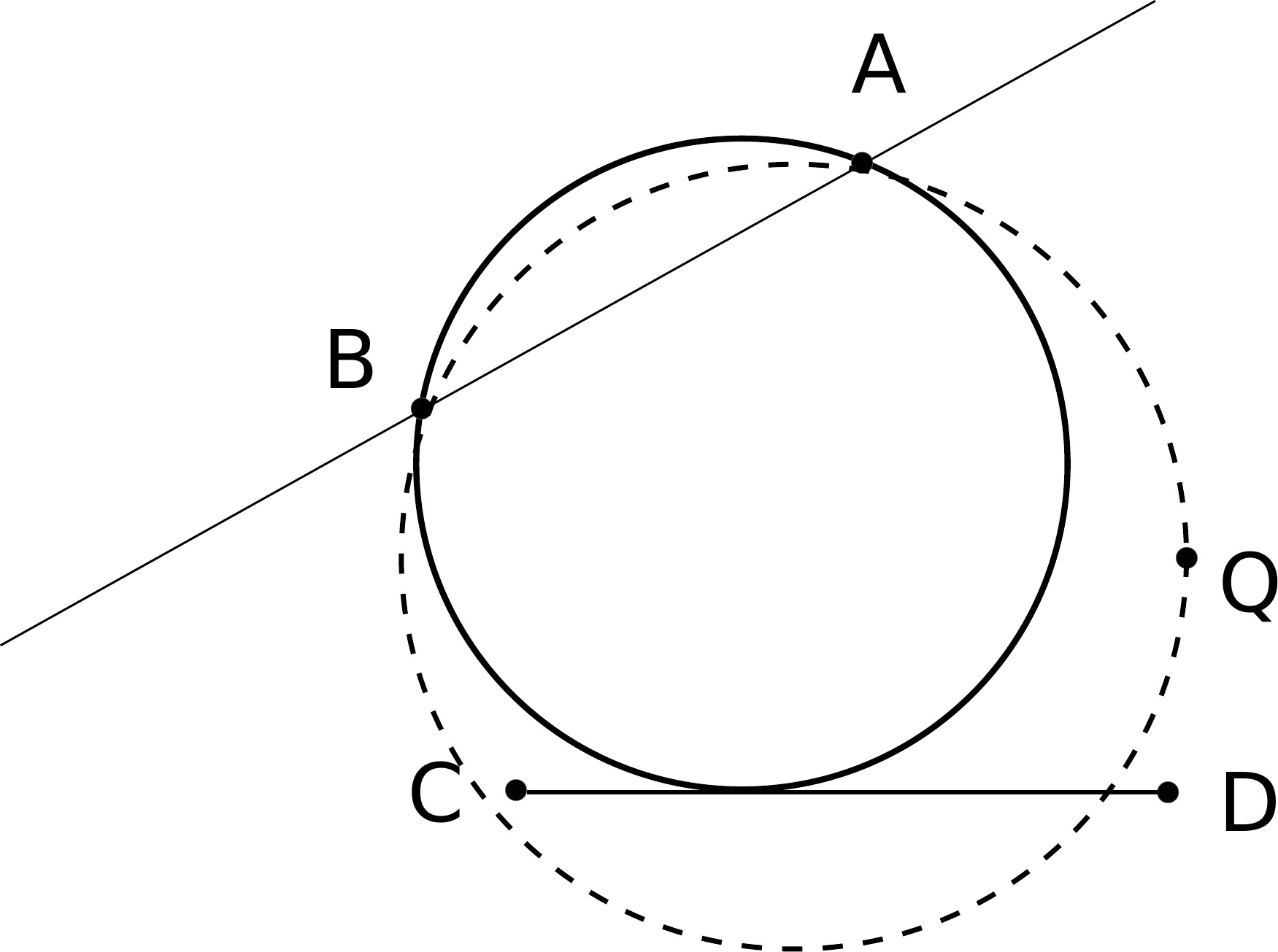}\\[3pt]%
    \includegraphics[width=0.4\textwidth]{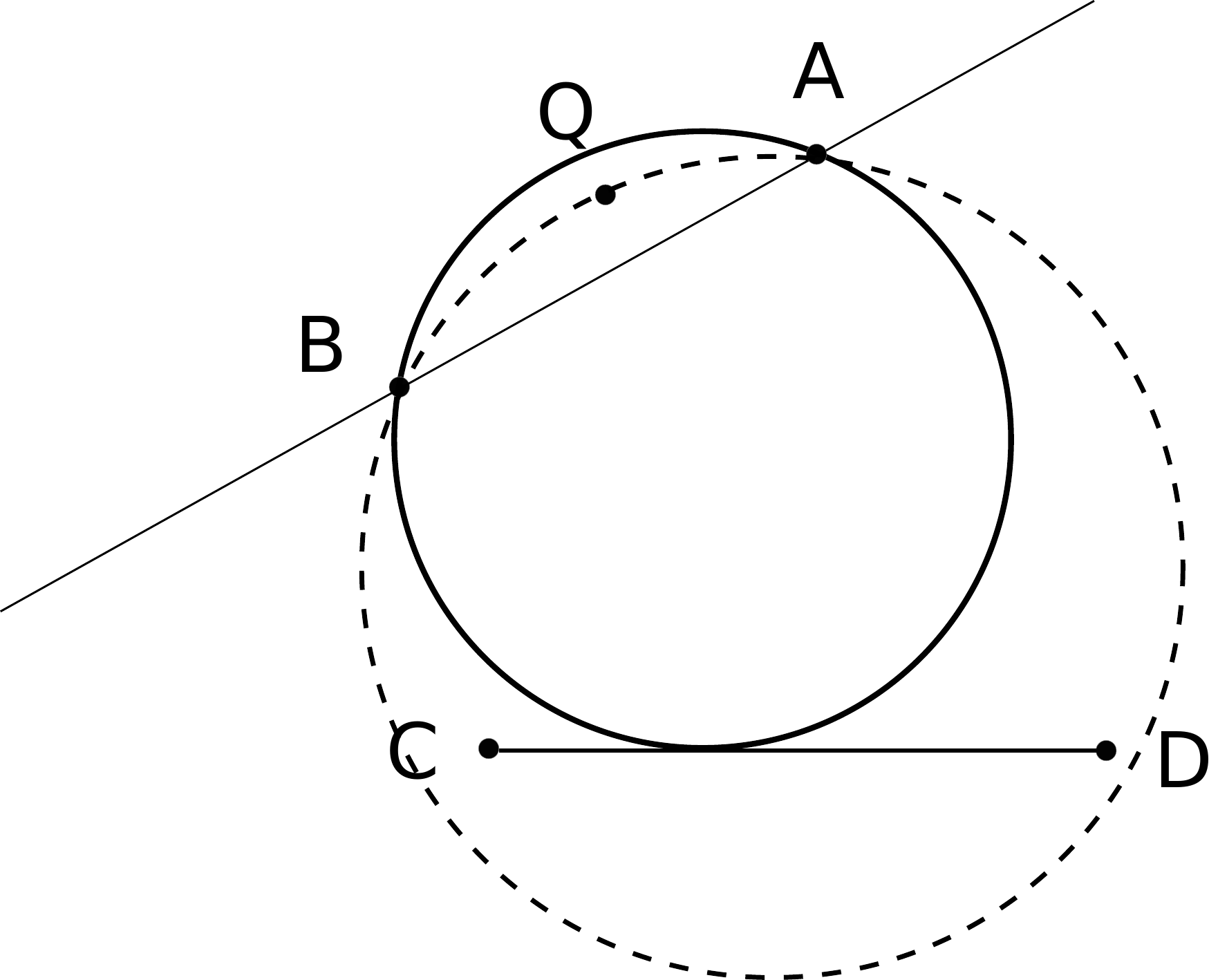}\hfil%
    \includegraphics[width=0.4\textwidth]{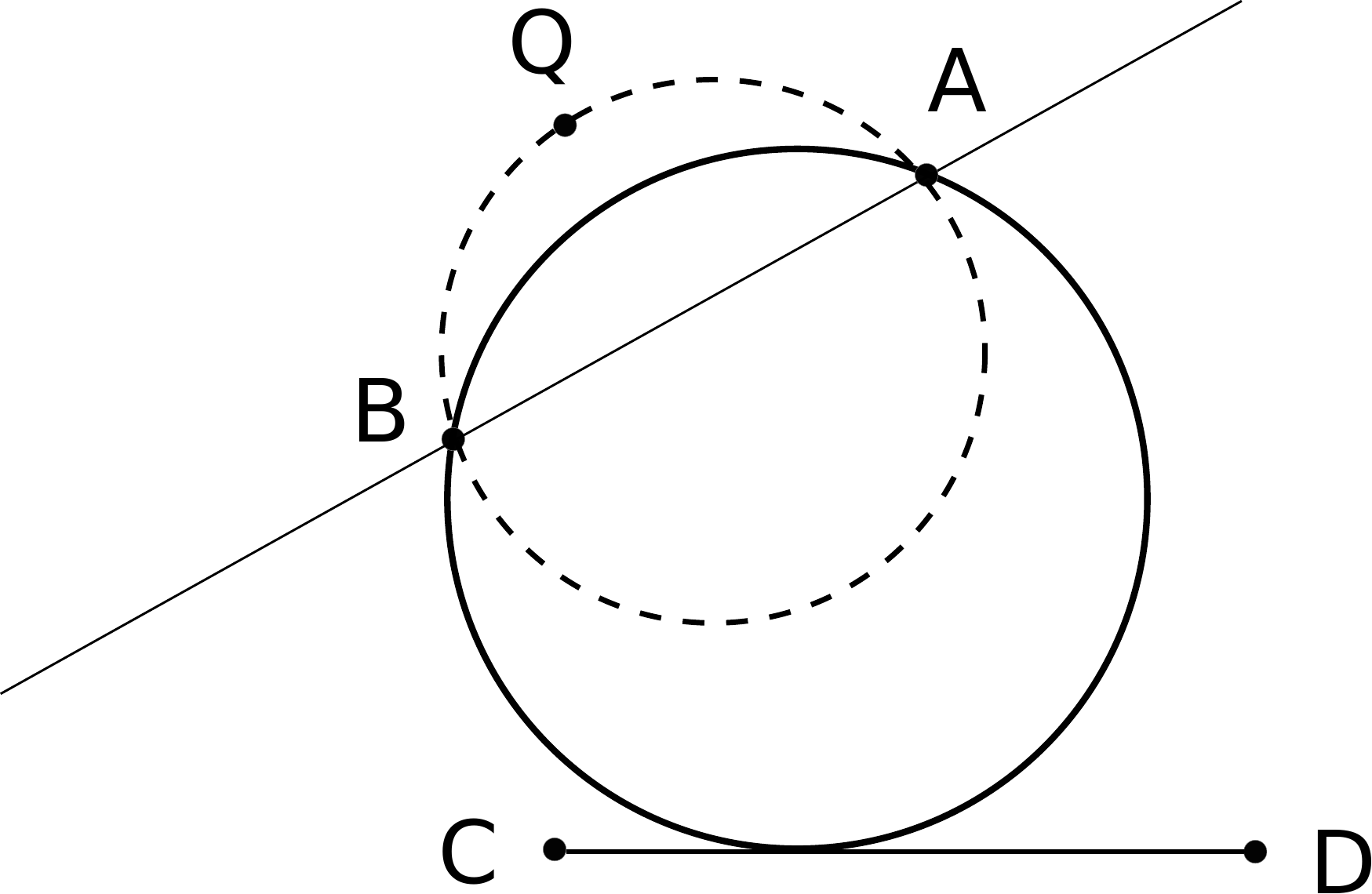}
  \end{center}
  \caption{Reducing $\incircle(A,B,CD,Q)$ to
    $\incircle(A,B,Q,CD)$. Top/Bottom row: $Q$ lies to the
    left/right of the oriented line $\ell_{AB}$.
    Left/Right column: $Q$ lies inside/outside $V(A,B,CD)$. The dotted
    circle is the \vor circle of $A$, $B$ and $Q$.}
  \label{fig:ppsp-reduction}
\end{figure}

Let $Q$ be the query point, and $K$ be the center of $V(A,B,CD)$. As
we will see in the next subsection, the $x$-coordinate of $K$ is a
root of a quadratic equation $P(x)=p_2x^2+p_1x+p_0$, where the
algebraic degrees of $p_2$, $p_1$ and $p_0$ are 1, 2 and 3,
respectively. Moreover, in this case
$y_K=\frac{\alpha_1}{\beta}x_K+\frac{\alpha_0}{\beta}$, where the
algebraic degrees of $\alpha_1$, $\alpha_0$ and $\beta$ are 1, 2 and
1, respectively (i.e., $\delta_\alpha=\delta_x=1$). By Subsection
\ref{sec:pxsp} we can evaluate $\incircle(A,B,CD,Q)$
using algebraic expressions of maximum degree
$2\cdot{}1+1+4=7$. Below, we are going to show
how to lower this maximum algebraic degree to 6.

Clearly, for the \vor circle $V(A,B,CD)$ to be defined, both $A$ and
$B$ must be on the same side with respect to $\ell_{CD}$. Consider now
$Q$: if $Q$ does not lie on the side of $\ell_{CD}$ that $A$ and $B$
lie, we have $\incircle(A,B,CD,Q)>0$. Testing the sideness of $I$,
$I\in\{A,B,Q\}$, against $\ell_{CD}$ simply means testing the sign of
$y_I-y_C$, which is a quantity of algebraic degree 1.

Suppose now that $Q$ lies on the same side of $\ell_{CD}$
as $A$ and $B$, and assume, without loss of generality, that
$\textsf{Orientation}(A,C,D)>0$ (the argument in the case
$\textsf{Orientation}(A,C,D)<0$, or when one of $A$ and $B$ lies on
$\ell_{CD}$, is analogous).
Consider the result $\sigma$ of the orientation
predicate $\textsf{Orientation}(A,B,Q)$. In the special case $\sigma=0$
(i.e., $Q$ lies on the line $\ell_{AB}$), we observe that $Q$ lies
inside the \vor circle $V(A,B,CD)$ if and only if $Q$ lies on
$\ell_{AB}$ and between $A$ and $B$. This can be determined by
evaluating the signs of differences $x_Q-x_A$ and $x_Q-x_B$, which are
both quantities of algebraic degree 1.

If $\sigma\ne{}0$, we are going to reduce $\incircle(A,B,CD,Q)$ to
$\incircle(A,B,Q,CD)$ (see also Fig. \ref{fig:ppsp-reduction}).
Suppose first that $\sigma>0$, i.e., $Q$ lies to the left of the oriented
line $\ell_{AB}$. Since $A$, $B$ and $CD$ appear on $V(A,B,CD)$ in
that order when we traverse it in the counterclockwise sense, we
conclude that $Q$ lies inside $V(A,B,CD)$ (resp., lies on $V(A,B,CD)$)
if and only if the circle defined by $A$, $B$ and $Q$,
does not intersect with (resp., touches) the line $\ell_{CD}$. To see
this, simply ``push'' the \vor circle towards $Q$, while keeping its
center on the bisector of $A$ and $B$. Hence,
$\incircle(A,B,CD,Q)=-\incircle(A,B,Q,CD)$.
In a similar manner, if $\sigma<0$, i.e., $Q$ lies to the right of the
oriented line $\ell_{AB}$, $Q$ lies inside $V(A,B,CD)$ (resp., lies on
$V(A,B,CD)$) if and only if the circle defined by $A$, $B$ and $Q$
intersects the line $\ell_{CD}$. Hence,
$\incircle(A,B,CD,Q)=\incircle(B,A,Q,CD)$.

Summarizing our analysis above, we first need to determine on which
side of $\ell_{CD}$ $Q$ lies: this a degree 1 predicate. If needed,
the next step is to compute $\textsf{Orientation}(A,B,Q)$, which is a
degree 2 predicate. If $\textsf{Orientation}(A,B,Q)=0$ we need two
additional tests of degree 1 to answer $\incircle(A,B,CD,Q)$;
otherwise, we observe that
\begin{equation*}
  \incircle(A,B,CD,Q)=\begin{cases}
    -\incircle(A,B,Q,CD),
    &\text{if\ }\textsf{Orientation}(A,B,Q)>0\\
    \textcolor{white}{-}\incircle(B,A,Q,CD),
    &\text{if\ }\textsf{Orientation}(A,B,Q)<0
  \end{cases}
\end{equation*}
As per Section \ref{sec:ppps}, $\incircle(A,B,Q,CD)$ or
$\incircle(B,A,Q,CD)$ can be answered using quantities of algebraic
degree at most 6. 


\subsection{The query object is a segment}\label{sec:ppss}

For this case we are going to follow the generic analysis presented in
Section \ref{sec:pxss}. Let $QS$ be the query segment, and
let $K$ be the center of $V(A,B,CD)$. $K$ is an intersection point 
of the bisector of $A$ and $B$ and the parabola with focal point $A$
and directrix the supporting line $\ell_{CD}$ of $CD$. Solving the
corresponding system of equations we deduce that, in the
general case where $A$ and $B$ are not equidistant from $\ell_{CD}$
(i.e., if $y_A\neq{}y_B$), the $x$-coordinate of the \vor center
$x_K$, is a root of the quadratic polynomial $P(x)=p_2x^2+p_1x+p_0$,
where
$p_{2}= y_{B}-y_{A}\neq{}0$,
$p_{1}= (y_{B}-y_{C}) (x_{A}-x_{B}) - 2x_{B}p_2$,
$p_{0}= p_2x_{B}^{2}+ (y_{C}-y_{B})[(x_{B}^{2}-x_{A}^{2})+(y_{A}-y_{C})p_2]$,
while the $y$-coordinate of the \vor center $y_K$, is a root of 
the quadratic polynomial $T(y)=t_2y^2+t_1y+t_0$, where
$t_2=4 (y_B-y_A)^2$,
$t_1=4(2y_C-y_A-y_B)(x_B-x_A)^2+4(y_B-y_A)(y_A^2-y_B^2)$,
$t_0=(x_A-x_B)^2(2y_A^2+2y_B^2 -4y_C^2 +(x_A-x_B)^2)+(y_A^2-y_B^2)^2$.
Moreover, $y_K$ and $ x_K $ are linearly dependent:
$y_K=\frac{\alpha_1}{\beta}x_K+\frac{\alpha_0}{\beta}$, where
$\alpha_1=2(x_A-x_B)$, $\alpha_0=x_B^2+y_B^2-x_A^2-y_A^2$ and 
$\beta=2(y_B-y_A)$.
The roots $x_1\leq{}x_2$ of the polynomial $P(x)$
(resp. $y_1\leq{}y_2$ of $T(y)$) correspond to the centers of the two
possible \vor circles $V(A,B,CD)$ and $V(B,A,CD)$.
The roots of $P(x)$ or of $T(y)$) of interest are shown in the
following two tables.
\begin{center}
\begin{tabular}{|c|c|}
\hline
Relative positions of $A$, $B$ and $CD$ & Root of $P(x)$ of interest\\
\hline \hline
 $ y_{C} < y_{A} < y_{B} $ & $x_{1}$\\ \hline
 $ y_{C} < y_{B} < y_{A} $ & $x_{2}$\\ \hline
 $ y_{B}< y_{A}<y_{C} $ & $x_{2}$\\ \hline
 $ y_{A} < y_{B} <y_{C} $& $x_{1}$\\ \hline
\end{tabular}\\[5pt]
\begin{tabular}{|c|c|}
\hline
Relative positions of $A$, $B$ & Root of $T(y)$ of interest\\
\hline \hline
$x_{A}<x_{B}$ & $y_{2}$ \\ \hline 
$x_{A}>x_{B}$ & $y_{1}$ \\ \hline 
\end{tabular}
\end{center}
The degrees of $p_2$, $p_1$, $p_0$, $t_2$, $t_1$ and $t_0$
are 1, 2, 3, 2, 3 and 4, respectively. Furthermore, the
degrees of $\alpha_1$, $\alpha_0$ and $\beta$ are 1, 2 and 1,
respectively. Applying the analysis in Subsection \ref{sec:pxss}
(where $\delta_\alpha=\delta_x=1$, $\delta_y=2$), we deduce that we
can answer the \incircle predicate using expressions of algebraic
maximum algebraic degree $\max\{2\cdot{}1+1+2,2+2\}=5$.

For the special case $y_A=y_B$, we easily get
$x_K=\frac{1}{2}(x_A+x_B)$ and $y_K=\frac{U_2}{U_1}$, where
$U_2 = (x_B-x_A)^2+4(y_A^2-y_C^2)$,
$U_1 = 8(y_A-y_C)$.
In this case, if $QS$ is $x$-axis parallel, we need to determine the
sign of the quantity
$d(K,\ell_{QS})-d(K,CD)=|y_K-y_Q|-|y_K-y_C|$, or, equivalently, the sign
of the quantity $|U_2-U_1y_Q|-|U_2-U_1y_C|$, which is of algebraic
degree 2. If $QS$ is $y$-axis parallel, we need to evaluate the sign
of the quantity $d(K,\ell_{QS})-d(K,CD)=|x_K-x_Q|-|y_K-y_C|$, or,
equivalently, the sign of the quantity
$|U_1(x_A+x_B-2x_Q)|-2|U_2-U_1y_Q|$, which is also of algebraic degree
2. 
%
%
Given, that the algebraic degree for the \ppsp case
is 6 (see previous subsection), 
we conclude that we can answer the \incircle predicate in the \ppss
case by computing the signs of expressions of algebraic degree at most
6.


\section{The \pss case}\label{sec:pss}

\subsection{The query object is a point}\label{sec:pssp}

In this section we consider the case where the \vor circle is defined
by two segments, a point and the query object is a point. Let $A$,
$CD$ and $FG$ be the point and the two segments defining the \vor
circle and let $Q$ be the query point. Since each of $CD$, $FG$ may be
$x$-axis or $y$-axis parallel we have four cases to consider:
(1) $CD$ and $FG$ are $x$-axis parallel,
(2) $CD$ and $FG$ are $y$-axis parallel,
(3) $CD$ is $x$-axis parallel and $FG$ is $y$-axis parallel, and
(4) $CD$ is $y$-axis parallel and $FG$ is $x$-axis parallel.
However, Cases (2) and (4) reduce to Cases (1) and (4), respectively,
by simply performing a reflection transformation through the line
$y=x$ (see Section \ref{sec:reflection}). More precisely, in both
cases we have
$\incircle(A,CD,FG,Q)=\incircle(\rfx{A},\rfx{FG},\rfx{CD},\rfx{Q})$. Thus,
for Case (2), $\rfx{CD}$ and $\rfx{FG}$ are $x$-axis parallel, while,
for Case (4), $\rfx{CD}$ is $x$-axis parallel and $\rfx{FG}$ is
$y$-axis parallel. Therefore it suffices to consider Cases (1) and (3).
In what follows, we follow the generic procedure described in
Subsection \ref{sec:pxsp}, and refer to the notation introduced
there.

\begin{figure}[t]
  \begin{center}
    \begin{minipage}{0.442\textwidth}
      \hspace*{1mm}\includegraphics[width=0.9\textwidth]{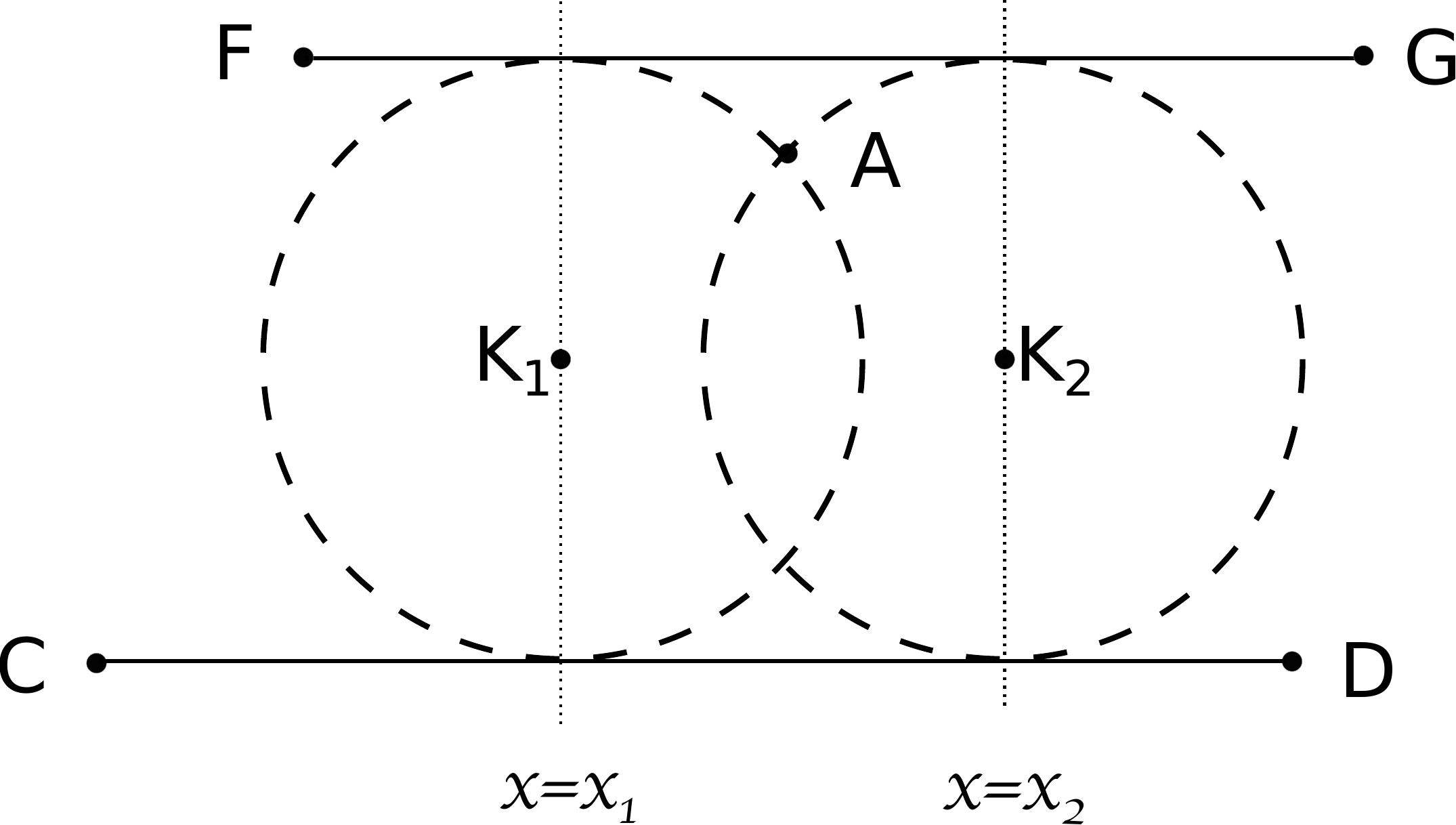}%
      \vspace*{2mm}
    \end{minipage}%
    \begin{minipage}{0.44\textwidth}
      \hspace*{5mm}\includegraphics[width=0.85\textwidth]{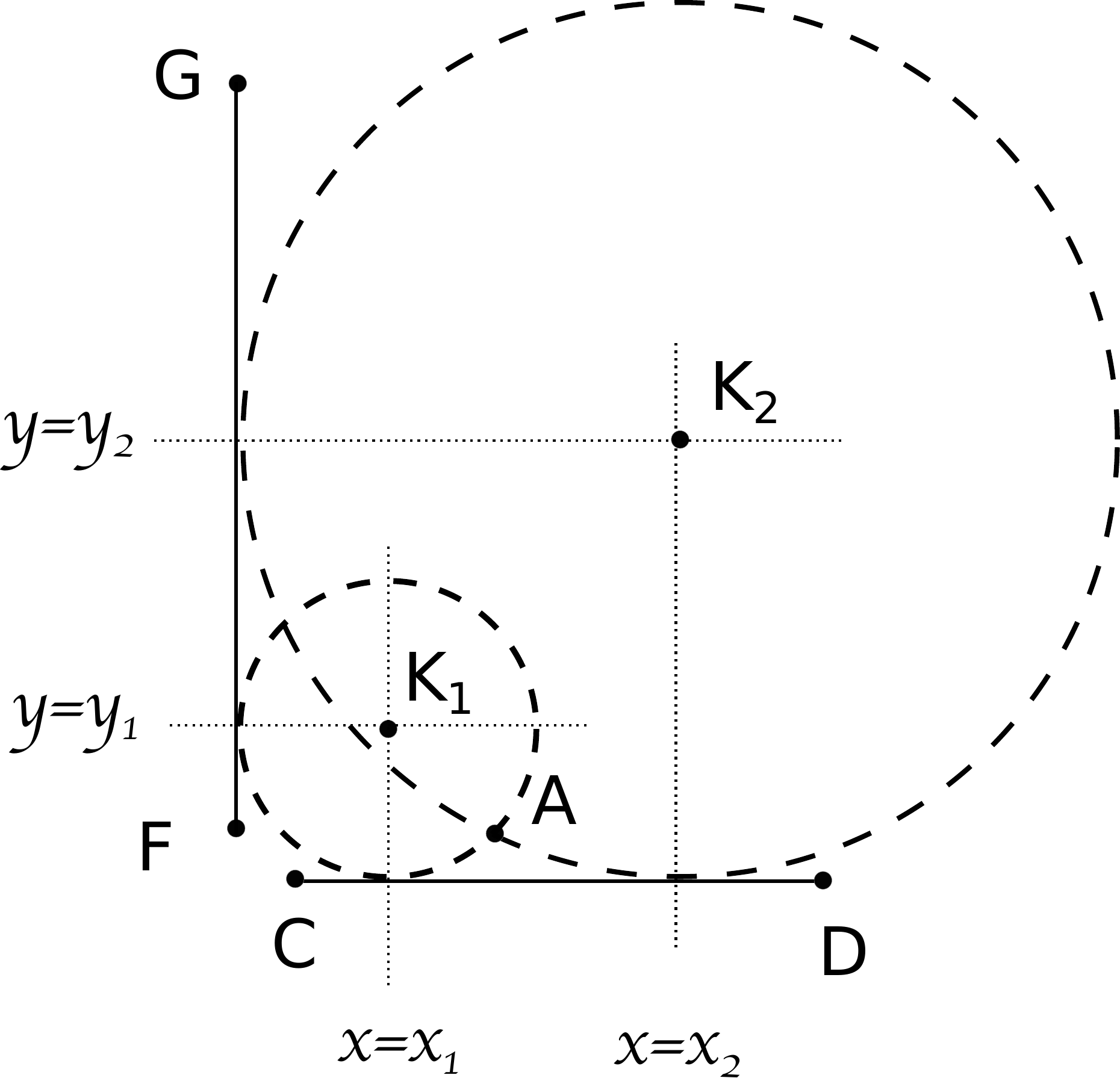}
    \end{minipage}
  \end{center}
  \caption{Voronoi circle defined by the point $A$ and the line
    segments $CD$ and $FG$. Left: $CD$, $FG$ are $x$-axis
    parallel. Right: $CD$ is $x$-axis parallel and $FG$ is $y$-axis
    parallel.}
  \label{fig:pss}
\end{figure}

\subsubsection{$CD$ and $FG$ are $x$-axis parallel}\label{sec:pssp-parallel}

We first notice that if $Q$ does not lie inside the band $B_x$
delimited by the $\ell_{CD}$ and $\ell_{FG}$, it cannot be inside the
\vor circle $V(A,CD,FG)$. This can be easily checked by evaluating the
signs of $y_Q-y_C$ and $y_Q-y_F$, which are quantities of algebraic
degree 1. Suppose now that $Q$ is inside $B_x$ and notice that $A$ has
to lie in $B_x$ in order for the \vor circle $V(A,CD,FG)$ to exist. 

Let $K$ be the center of $V(A,CD,FG)$. The $y$-coordinate of $K$ is,
trivially, $y_K=\frac{1}{2}(y_C+y_F)$, whereas the radius $\rho$ of
the \vor circle is equal to $\rho=\frac{1}{2}|y_C-y_F|$. Given that
$A$ is a point on $V(A,CD,FG)$, we have that $d^2(K,A)=\rho^2$. Using
the expressions for $y_K$ and $\rho$, we deduce that $x_K$ is a root
of the polynomial $P(x)=x^2+p_1x+p_0$, where
$p_{1}=2x_A$ and $p_{0}=x_A^{2}+(y_{A}-y_{C})(y_{A}-y_{F})$.
If $x_1\le{}x_2$ are the two roots of $P(x)$, the root that
corresponds to $x_K$ is given in the table below (see also
Fig. \ref{fig:pss}(left)).
\begin{center}
\begin{tabular}{|c|c|}
\hline
Relative positions of $A$ and $CD$ & Root of $ P(x) $ of interest\\
\hline \hline
$y_A>y_C$&$x_{2}$\\\hline
$y_A<y_C$&$x_{1}$\\\hline
\end{tabular}
\end{center}
Moreover, in this case we have $\alpha_1=0$,
$\alpha_0=y_C+y_F$ and $\beta=2$. Therefore, the algebraic degrees
involved in the evaluation of the \incircle predicate are
$\delta_\alpha=\delta_x=0$. As per Subsection \ref{sec:pxsp},
the $\incircle(A,CD,FG,Q)$ predicate can be evaluated using algebraic
expressions of maximum degree $2\cdot{}0+0+4=4$.

\subsubsection{$CD$ is $x$-axis parallel and $FG$ is $y$-axis parallel}
\label{sec:pssp-vertical}

The lines $\ell_{CD}$ and $\ell_{FG}$ subdivide the plane into
four quadrants $R_1$, $R_2$, $R_3$ and $R_4$.
The bisector of $R_1$ and $R_3$ is the line
$\ell_{1,3}$ with equation $y=x+y_C-x_F$, whereas the bisector of
$R_2$ and $R_4$ is the line $\ell_{2,4}$ with equation $y=-x+y_C+x_F$.

The center $K$ of the \vor circle $V(A,CD,FG)$ lies on both the
bisector of $\ell_{CD}$ and $\ell_{FG}$, as well as on the parabola
that is at equal distance from $A$ and $\ell_{CD}$; the equation of
the latter is:
\begin{equation}\label{equ:parabola-A-CD}
  (x-x_A)^2-(y_A-y_C)(2y-y_A-y_C)=0.
\end{equation}
Assuming that $A$ lies in $R_1\cup{}R_3$, the bisector of $\ell_{CD}$
and $\ell_{FG}$ is $\ell_{1,3}$. Substituting $y$ in terms of $x$,
using the equation of $\ell_{1,3}$, we deduce that the $x$-coordinate
$x_K$ of $K$ is a root of the quadratic polynomial
$P(x)=x^2+p_1x+p_0$, where
$p_1=2(y_C-y_A-x_A)$, and
$p_0=(y_C-y_A)^2+x_A^2-2x_F(y_C-y_A)$.
Similarly, if $A$ lies in $R_2\cup{}R_4$, $x_K$ is a root of the
quadratic polynomial $P(x)=x^2+p_1x+p_0$, where
$p_1=2(y_A-y_C-x_A)$,
$p_0=(y_C-y_A)^2+x_A^2+2x_F(y_C-y_A)$.
If $x_1\le{}x_2$ are the two roots of $P(x)$, the root that
corresponds to $x_K$ is the same as in the case where $FG$ is $x$-axis
parallel.
Moreover, in this case we have $\alpha_1=1$,
$\alpha_0=y_C-x_F$, $\beta=1$, if $A\in{}R_1\cup{}R_3$, and
$\alpha_1=-1$, $\alpha_0=y_C+x_F$, $\beta=1$, if
$A\in{}R_2\cup{}R_4$. In both cases, the algebraic degrees
involved in the evaluation of the \incircle predicate are
$\delta_\alpha=\delta_x=0$. Again, as per Subsection \ref{sec:pxsp},
the $\incircle(A,CD,FG,Q)$ predicate can be evaluated using algebraic
expressions of maximum degree $2\cdot{}0+0+4=4$.



\subsection{The query object is a segment}\label{sec:psss}

Let $QS$ be the query segment, while the \vor
circle is defined by the point $A$ and the segments $CD$ and $FG$. Let
$K=(x_K,y_K)$ be the center of the \vor circle. As in the previous
subsection, it suffices to consider the cases where, either both $CD$
and $FG$ are $x$-axis parallel, or $CD$ is $x$-axis parallel and $FG$
is $y$-axis parallel. Recall that, in both cases, we have shown that
$x_K$ is always a root of a quadratic polynomial $P(x)=x^2+p_1x+p_0$,
where the algebraic degrees of $p_1$ and $p_0$ are 1 and 2, respectively.

\subsubsection{$CD$ and $FG$ are $x$-axis parallel}\label{sec:psss-parallel}

If $QS$ is also $x$-axis parallel we first need to determine if $QS$
lies inside the band $B_x$ delimited by $\ell_{CD}$ and
$\ell_{FG}$. This is easily done by checking if $Q$ lies inside $B_x$,
which in turn means checking the signs of $y_Q-y_C$ and $y_Q-y_F$, as
described in the previous subsection. Clearly, if $Q$ is not inside
the band $B_x$, then $\incircle(A,CD,FG,QS)>0$.
Assume now that $QS$ lies inside $B_x$. The first step is to evaluate
the $\incircle(A,CD,FG,Q)$ and, if necessary,
$\incircle(A,CD,FG,S)$. If $\incircle(A,CD,FG,Q)<0$ or
$\incircle(A,CD,FG,S)<0$, then we immediately know that
$\incircle(A,CD,FG,QS)<0$. Otherwise, we simply need to determine on
which side of the line $x=x_K$ $Q$ and $S$ lie: $QS$ intersects the
\vor circle $V(A,CD,FG)$ if and only if $Q$ and $S$ lie on different
sides of $x=x_K$. Determining the side of $x=x_K$ on which the point
$I$, $I\in\{Q,S\}$, lies is equivalent to computing the sign of the
difference $x_K-x_I$. This, in turn, reduces to computing the
signs of the expressions $P(x_I)$ and $P'(x_I)$, which are
expressions of algebraic degree 2 and 1, respectively.

In the case where $QS$ is $y$-axis parallel, we proceed according to
the generic approach presented in Subsection \ref{sec:pxss}. In this
case $y_K=\frac{1}{2}(y_C+y_F)$, i.e., $\alpha_1=0$,
$\alpha_0=y_C+y_F$ and $\beta=2$. Moreover, $T(y)$ is a linear
polynomial $T(y)=2y-(y_C+y_F)$, thus the algebraic degrees of $T(y_I)$
and $T'(y_I)$, $I\in\{Q,S\}$, are $\delta_y+1=1$ and $\delta_y=0$,
respectively. By applying the analysis of Subsection \ref{sec:pxss},
with $\delta_x=\delta_\alpha=\delta_y=0$, 
we conclude that we can answer the \incircle predicate by evaluating
the signs of expressions of algebraic degree at most
$\max\{2\cdot{}0+0+4,0+1\}=4$.

\subsubsection{$CD$ is $x$-axis parallel and $FG$ is $y$-axis
  parallel}\label{sec:psss-vertical}

For the purposes of resolving this case, we are going to follow the
analysis of Subsection \ref{sec:pxss}. In the previous subsection
we argued that in this case the center $K=(x_K,y_K)$ of the \vor
circle $V(A,CD,FG)$ lies on the intersection of the parabola with
equation \eqref{equ:parabola-A-CD} and either the line $y=x+y_C-x_F$
(if $A\in{}R_1\cup{}R_3$) or the line $y=-x+y_C+x_F$ (if
$A\in{}R_2\cup{}R_4$). Solving in terms of $y$ we deduce that $y_K$ is
a root of the quadratic polynomial $T(y)=y^2+t_1y+t_0$, where
$t_1= -2 (y_A+x_A+x_F-2y_C)$,
$t_0= (x_A+x_F)^2 + y_A^2-2y_C(x_A+x_F)$,
if $A\in{}R_1\cup{}R_3$, whereas
$t_1= 2(x_A-y_A-x_F)$,
$t_0= (x_A-x_F)^2+ y_A^2-2x_Ay_C+2y_Cx_F$,
if $A\in{}R_2\cup{}R_4$.
Notice that in both cases the algebraic degrees of $t_1$ and $t_0$ are
1 and 2, respectively. Furthermore, if $y_1\le{}y_2$ are the two roots
of $T(y)$, the root of $T(y)$ of interest is given in the following
table (see also Fig. \ref{fig:pss}(right)).
\begin{center}
\begin{tabular}{|c|c|}
\hline
Relative positions of $A$ and $FG$ & Root of $ T(y) $ of interest\\
\hline \hline
$x_A>x_F$&$y_{2}$\\\hline
$x_A<x_F$&$y_{1}$\\\hline
\end{tabular}
\end{center}
Finally, as already 
described in the previous subsection, in this case we have $\alpha_1=1$,
$\alpha_0=y_C-x_F$, $\beta=1$, if $A\in{}R_1\cup{}R_3$, and
$\alpha_1=-1$, $\alpha_0=y_C+x_F$, $\beta=1$, if $A\in{}R_2\cup{}R_4$.
We are now ready to apply the analysis of Subsection \ref{sec:pxss},
with $\delta_\alpha=\delta_x=\delta_y=0$. We thus conclude that 
the predicate $\incircle(A,CD,FG,QS)$ can be evaluated using algebraic
quantities of degree at most $\max\{2\cdot{}0+0+4,0+2\}=4$.

\section{Conclusion and future work}\label{sec:concl}

In this paper we have studied the \incircle predicate involved in the
computation of the Euclidean \vor diagram for axes-aligned line
segments. We have described in detail, and in a self-contained manner,
how to evaluate this predicate. We have shown that we can always
resolve it using polynomial expressions in the input quantities that
are of maximum algebraic degree 6.

Our analysis is thus far theoretical. We would like to implement the
approach presented in this paper and compare it against the generic
implementation in CGAL \cite{cgal:k-sdg2-10b}.
Finally, we would like to study 
the rest of the predicates involved in the computation of the \vor
diagram, as well as consider the ortho-45$^\circ$ case, i.e., the case
where the segments are allowed to lie on lines parallel to the lines
$y=x$ and $y=-x$.


\section*{Acknowledgments}
\noindent
Work partially supported by the FP7-REGPOT-2009-1 project ``Archimedes
Center for Modeling, Analysis and Computation''.

\bibliographystyle{plain}
\bibliography{biblio,svd-vd04}

\end{document}